\documentclass[mathleft]{an}
\usepackage{graphicx}
\usepackage{times}
\begin{document}

\Pagespan{789}{}
\Yearpublication{2006}%
\Yearsubmission{2005}%
\Month{11}%
\Volume{999}%
\Issue{88}%

\title{Asteroseismology of Solar-type stars with Kepler III. Ground-based Data}

\author{J.\ Molenda-\.Zakowicz\inst{1}\fnmsep\thanks{\email{molenda@astro.uni.wroc.pl}\newline}, 
H.\ Bruntt\inst{2}, S.\ Sousa\inst{3},  A.\ Frasca\inst{4}, K.\ Biazzo\inst{5}, 
D.\ Huber\inst{6}, M.\ Ireland\inst{6}, T.\ Bedding\inst{6}, D.\ Stello\inst{6}, K.\ Uytterhoeven\inst{7},
S.\ Dreizler\inst{8}, P.\ De Cat\inst{9}, M.\ Briquet\inst{10}, G.\ Catanzaro\inst{4},
C.\ Karoff\inst{11,12}, S.\ Frandsen\inst{12}, L.\ Spezzi\inst{13}, and C.\ Catala\inst{2}
}
\titlerunning{Characteristics of 100+ Kepler Asteroseismic Targets from Ground-Based
Observations}
\authorrunning{J.\ Molenda-\.Zakowicz et al.}
\institute{
Instytut Astronomiczny Universytetu Wroc\l{}awskiego, ul.\ Kopernika 11, 51-622 Wroc\l{}aw, Poland
\and
LESIA, UMR\,8109, Universit\'e Pierre et Marie Curie, Universit\'e Denis Diderot, Observatoire de Paris, 92195 Meudon, France
\and
Centro de Astrof\'isica, Universidade do Porto, Rua das Estrelas, 4150-762, Portugal 
\and 
INAF--Osservatorio Astrofisico di Catania, Via Sofia 78, 95123 Catania, Italy
\and 
INAF--Osservatorio Astrofisico di Arcetri, Largo E. Fermi 5, 50125 Firenze, Italy
\and
Sydney Institute for Astronomy (SIfA), School of Physics, University of Sydney, NSW 2006, Australia 
\and
Laboratoire AIM, CEA/DSM-CNRS-Université Paris Diderot-IRFU/SAp, 91191 Gif-sur-Yvette Cedex, France
\and
Georg-August Universit\"at, Institut f\"ur Astrophysik, Friedrich-Hund-Platz 1, D-37077 G\"ottingen, Germany 
\and
Royal Observatory of Belgium, Ringlaan 3, B-1180 Brussels, Belgium
\and
Instituut voor Sterrenkunde, KU Leuven, Celestijnenlaan 200D, B-3001 Leuven, Belgium
\and
School of Physics and Astronomy, University of Birmingham, Edgbaston, Birmingham B15 2TT, UK 
\and
Department of Physics and Astronomy, Aarhus University, Ny Munkegade 120, DK-8000 Aarhus C, Denmark
\and
European Space Agency (ESTEC), P.O. Box 299, 2200 AG Noordwijk, The Netherlands
}

\received{1 April 2010}
\accepted{---}
\publonline{later}


\abstract{%
We report on the ground-based follow-up program of spectroscopic and photometric observations 
of solar-like asteroseismic targets for the Kepler space mission. These stars constitute a 
large group of more than thousand objects which are the subject of an intensive study of the Kepler 
Asteroseismic Science Consortium Working Group 1 (KASC WG-1). The main goal of this coordinated 
research is the determination of the fundamental stellar atmospheric parameters, which are used 
for the computing of their asteroseismic models, as well as for the verification of the Kepler
Input Catalogue (KIC).}

\maketitle

\section{Introduction}
The Kepler Asteroseismic Science Consortium KASC\footnote{Kepler
Asteroseismic Science Consortium (KASC) is a group of collaborating 
scientists established to accomplish the activities 
of the Kepler Asteroseismic Investigation (KAI), represented by Ronald Gilliland
(see http://astro.phys.au.dk/KASC).} has been established with the aim of making a full
exploitation of the Kepler time-series space photometry. The main goal of KASC is the 
study of stellar interiors by means of asteroseismic methods.

The largest group of stars selected for asteroseismic targets for Kepler are those which
are expected to show solar-like oscillations. Unfortunately, the Kepler data do not provide 
information on the effective temperature $(T_{\rm eff})$ surface gra\-vi\-ty, $(\log g)$, 
metallicity, and projected rotational velocity, $(v\sin i)$, of these stars, which are crucial 
data for asteroseismic modeling. Only a small fraction of Kepler asteroseismic targets 
has been studied in the literature. The effective temperature has been derived either from spectroscopy 
or photometry only for 15\% of the solar-like Kepler asteroseismic targets, the surface gravity, for 10\%, 
and 5\% have a measured metallicity. 

The Kepler Input Catalog\footnote{http://archive.stsci.edu/kepler/kepler\_fov/search.php} (KIC)
(Latham et al.\ 2005) provides values of $T_{\rm eff}$, $\log g$ and [Fe/H] derived from
Sloan photometry for many Kepler targets. However,  
the uncertainties of $T_{\rm eff}$, $\log g$ and [Fe/H] in the KIC are too large for asteroseismic modeling 
since they reach $\pm 200$\,K in $T_{\rm eff}$, and $\pm 0.5$\,dex, both in $\log g$ and [Fe/H]. 

\begin{figure}
\includegraphics[height=80mm,angle=270]{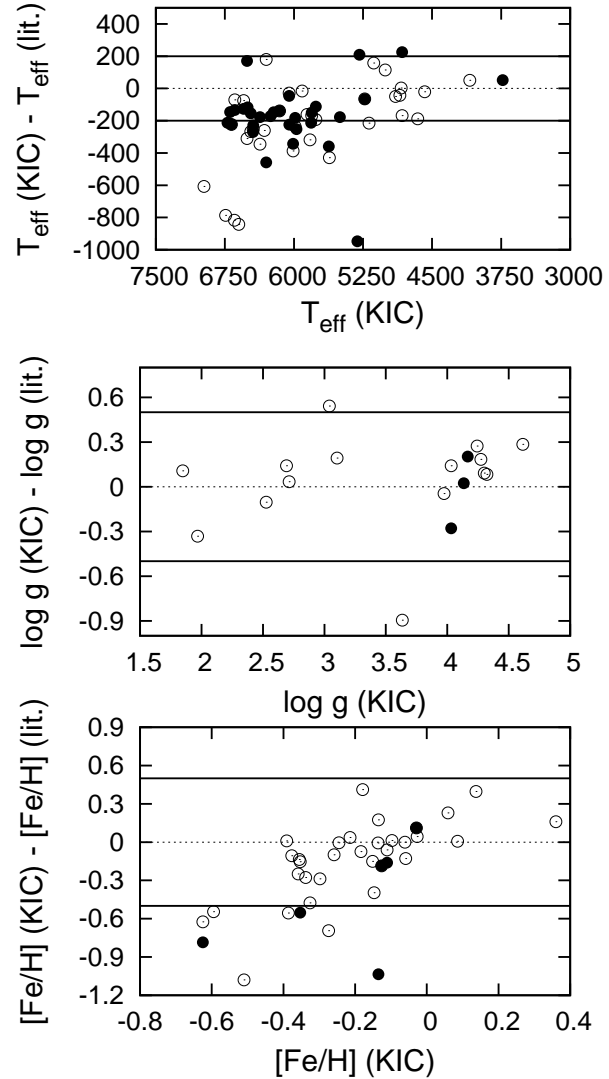}
\caption{The differences between $T_{\rm eff}$, $\log g$ and [Fe/H] given in the KIC and in the literature.
Solid and open symbols indicate literature values from photometry and spectroscopy, respectively.
The $1\sigma$ uncertainty ranges of the determinations provided in the KIC are indicated with solid lines.
}
\label{kic}
\end{figure}

In Fig.\ \ref{kic} we show the differences between $T_{\rm eff}$, $\log g$ and [Fe/H] of the
solar-like oscillators in the KIC and in the literature. As can be seen from the figure, 
the discrepancies can be as high as 1000\,K, especially for stars hotter than 5500\,K. 
The agreement in $\log g$ and [Fe/H] is slightly better but still poor. Their values 
agree to within the error bars, but the latter are so large that they do not allow any meaningful comparison.

Moreover, we note that there is a trend in the differences between [Fe/H] in the KIC and in the literature,
which exists both for the spectroscopic and photometric derivations, and which shows that the 
metal-deficient stars are found to be more metal-poor, and the metal-rich stars, more metal-abundant in 
the KIC than in the literature. 

\section{Programmes of Ground-Based Observations}

For the reasons outlined in the previous section, the KASC WG-1 Sub-Group 9 (SG-9) organized a large coordinated 
programme of ground-based observations which aims at the determination of $T_{\rm eff}$, $\log g$, [Fe/H], 
and $v\sin i$ of all the solar-like asteroseismic targets. This includes multi-co\-lour pho\-toelectric CCD 
photometry, high and me\-dium-re\-so\-lu\-tion spectroscopic observations, and interferometric measurements.

\begin{table*}
\centering
\caption{A summary of the programmes of ground-based observations of solar-like Kepler asteroseismic 
targets.}
\label{programs}
\begin{tabular}{cccrc}\hline
\noalign{\smallskip}
Observatory & Telescope/ & type of    & number of & P.I.\\
            & Instrument & programme  & nights~~~~&\\
\noalign{\smallskip}
\hline
\noalign{\smallskip}
Osservatorio Astrofisico di Catania, Italy               & 91-cm FRESCO    & spectroscopy& 92 & J.M-\.Z\\
$\prime \prime$                                          & 91-cm photometer& photometry & 11 & J.M-\.Z\\
Observatorio del Roque de los Muchachos, La Palma, Spain & 2.54-m INT WFC  & photometry  &  5 & K.U.\\
$\prime \prime$                                          & 2.56-m NOT FIES & spectroscopy&  3 & K.U.\\
$\prime \prime$                                          & $\prime \prime$ & spectroscopy&  7 & S.F.\\
$\prime \prime$                                          & 1.2-m Mercator HERMES& spectroscopy&  7 & M.B.\\
$\prime \prime$                                          & 3.58-m TNG SARG & spectroscopy& 12 & G.C.\\
Calar Alto, Spain                                        & 2.2-m BUSCA     & photometry  &  5 & K.U.\\
Observatory of Iza\~na, Tenerife, Spain                  & 0.8-m IAC-80 CAMELOT & photometry  & 14 & K.U.\\
Mauna Kea, USA                                           & 3.6-m CFHT ESPaDOnS  & spectroscopy& 10 hrs & H.B.\\
Pic du Midi, France                                      & 2-m NARVAL      & spectroscopy& 40 hrs & H.B.\\
Xinglong, China (proposal submitted, under consideration)& 4-m LAMOST      & spectroscopy& 13  & P.D.C.\\
Center for High Angular Resolution Astronomy, USA        & CHARA PAVO      & interferometry& 6 & D.H. \\
\hline
\end{tabular}
\end{table*}

In Table \ref{programs}, we list the on-going programmes of ground-based observations of solar-like 
Kepler asteroseismic targets, including the information on the observatory at which the programmes are 
realized, the telescope and the instrument used, the type of the observing programme (spec\-tro\-sco\-py, 
photometry or interferometry), the number of the ni\-ghts, and the initials of the principal investigator of 
the proposal. 

We note that many of the listed research programmes include targets of 
different KASC working groups. For the additional information, we refer to Uytterhoeven et al.\ (2010).

\subsection{Spectroscopy}

The spectroscopic investigation of the Kepler asteroseismic targets aims at derivation of the atmospheric 
parameters of the stars and the value of the micro-turbulence in their atmospheres, as well as the 
projected rotational velocity and the radial velocity (see Molenda-\.Zakowicz et al. 2007, 
2008, 2010; Catanzaro et al. 2010; Frasca et al. 2010.) 

The observatories involved in this research include the Osservatorio Astrofisico di Catania (OACt) in Italy, 
the Observatorio del Roque de los Muchachos in Spain, the Mauna Kea Observatory in the USA, 
the Pic du Midi Observatory in France, and the Xinglong Observatory in China. We note that the observing proposal 
submitted by P.D.C.\ for the multi-fiber, multi-object spectrograph LAMOST at the 4-m telescope at Xinglong is 
particularly interesting because, if accepted, it will allow to acquire low and/or medium-resolution spectra of 95\% 
of all the Kepler asteroseismic targets.

A separate spectroscopic investigation of late-type main sequence stars selected for being observed by Kepler for 
the entire life-time of the mission is carried out by C.K.\ (Karoff et al.\ 2009). The aim of the research of these 
authors is to understand the relation between the changes in the chromospheric activity of the stars and the changes 
in the eigenmodes of their oscillations, which can lead to the improvement of the models of stellar evolution.

An investigation of giant stars from the Kepler field of view is conducted by S.F., D.S. and H.B. who have been
awarded time to acquire high resolution spectra (R=60\,000) of these stars with FIES at the NOT. The aim is to get
a signal-to-noise ratio (S/N) of 100 for about 60-80 giant stars. C.C and H.B. have been awarded time to observe 300
solar-type stars at Pic du Midi and CFHT. Their aim is to get S/N = 100-200 with a resolution of R=80\,000.
With these data, they will be able to derive $T_{\rm eff}$, $\log g$, and [Fe/H] of the targets with a 1\,$\sigma$
uncertainty of 100K in Teff, and 0.1 dex in both log g and [Fe/H]. 

\subsection{Photometry}
The photometric research realized at the OACt in Italy, the Observatory of Iza\~na and Calar Alto Observatory 
in Spain also allows to determine the stellar atmospheric parameters from the Johnson and Str\"omgren magnitudes. 
These data, when combined with 2MASS JHK magnitudes (Cutri et al.\ 2003), can be used for computing the spectral 
energy distribution of the stars and deriving their atmospheric parameters. Additionally, photometric observations 
allow to measure the reddening of the stars and their absolute magnitude (see Molenda-\.Zakowicz, Jerzykiewicz \& 
Frasca 2009; Frasca et al.\ 2009; Kupka \& Bruntt 2001.) 

\subsection{Interferometry}

The interferometric measurements performed by M.I., D.H., T.B., and D.S. using the Center for High Angular Resolution 
Astronomy (CHARA, ten Brummelaar et al.\ 2005) array at Mt. Wilson Observatory, California, USA, aim at deriving the 
angular diameters, $\theta$, of the brightest Kepler stars (V=7-9 mag). Observations are performed at visible 
wavelengths using the Precision Astronomical Visible Observations (PAVO) beam combiner (Ireland et al.\ 2008). 

Using the longest CHARA baselines (330m) at a central wavelength of 0.75 $\mu$m, PAVO is able to resolve stars as small 
as 0.3 mas, which will allow precise angular diameter measurements of main-sequence stars observed by Kepler. 
Combining $\theta$ with the parallax yields a measurement of the linear radius, while a combination of $\theta$ with 
the bolometric flux yields a direct measurement of the effective temperature. 

\begin{figure}
\includegraphics[width=80mm]{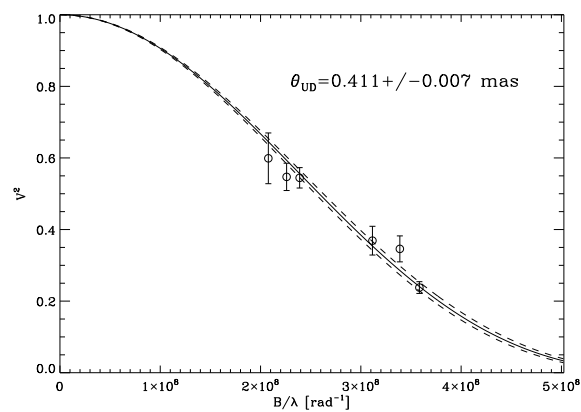}
\caption{Squared visibility versus spatial frequency of a metal-poor sub-giant in the Kepler target list obtained 
using the PAVO beam combiner at the CHARA array. The solid line shows a uniform-disc diameter fit with 1-sigma 
uncertainties (dashed lines).}
\label{inter}
\end{figure}

Fig.\ \ref{inter} shows an example of preliminary PAVO data obtained in July 2009 (Huber et al.\ in preparation) 
of a metal-poor sub-giant observed by Kepler. 

\section{Atmospheric parameters}

The atmospheric parameters of solar-like Kepler asteroseismic targets are the most important result of
the research conducted by KASC WG-1 SG-9 since their values are crucial for asteroseismic modeling. In order 
to find precise values of these parameters and to estimate their external errors we use different analysis methods
applied to the same data, as well as to the data acquired by various instruments or by different 
methods, i.e., spectroscopic or photometric.

For the derivation of the atmospheric parameters of stars from spectroscopic observations, we use the 
VWA\footnote{http://physics.usyd.edu.au/~bruntt/vwa/} software developed by H.B.\ (see Bruntt et al.\ 2004, 2010), 
the ARES\footnote{http://www.astro.up.pt/~sousasag/ares/} software developed by S.S.\ (Sousa et al.\ 2007), 
the MOOG code developed by Sneden (1973)\footnote{http://verdi.as.utexas.edu/}, the ROTFIT code developed by 
A.F.\ (Frasca et al.\ 2003, 2006), and the TLUSTY/SYNSPEC developed by Hubeny (1995).

As shown by Metcalfe et al.\ (in preparation) who discuss KIC 11026764 observed at the NOT with the FIES instrument, 
the atmospheric parameters derived by means of the methods listed above agree well to within 1$\sigma$ of 
their error bars. Nevertheless, the external uncertainty of the derivation of $T_{\rm eff}$ remains $\pm200$\,K 
while for $\log g$ and [Fe/H] it has been reduced to $\pm$0.25 and $\pm$0.1 dex, respectively.

Another kind of disagreement in $T_{\rm eff}$ derived from spectroscopy and from Str\"omgren photometry has been 
noticed by Molenda-\.Zakowicz et al.\ (2009) who show that stars from the temperture range 6500 -- 7500\,K in photometry 
are found systematically cooler by about 300\,K in spectroscopy. This effect is of particular significance since it 
affects the process of the asteroseismic modeling of the stars, and as such should be explained. This will be done when 
more spectroscopic and photometric observations of Kepler asteroseismic targets are acquired.

\section{Future observations}

In the future, we will not restrict the ground-based spectroscopic and photometric follow-up observations to large 
telescopes, but we intend to additionally include small and medium-size telescopes because many Kepler targets that are 
bright enough for such instruments. Additionally, at the sites hosting these instruments it is often possible to perform 
long and dedicated observing runs that our observing program requires.

For these reasons, we strongly encourage the astronomical community to use the small and medium-size Northern 
telescopes for the photometric and spectroscopic research of Kepler asteroseismic targets. We stress that 
the participation in this programme will be beneficial also for the sites hosting small telescopes.

\acknowledgements
This work was supported by MNiSW grant N203 014 31/2650 and the University of 
Wroc{\l}aw grants No 2646 /W/IA/06 and 2793/W/IA/07.
M.B.\ is a Postdoctoral Fellow of the Fund for Scientific Research, Flanders.

\end{document}